\documentclass[a4paper,fleqn]{cas-sc}
\usepackage{soul}
\usepackage[sort&compress,numbers]{natbib}
\usepackage{setspace}
\def\tsc#1{\csdef{#1}{\textsc{\lowercase{#1}}\xspace}}
\tsc{WGM}
\tsc{QE}
\tsc{EP}
\tsc{PMS}
\tsc{BEC}
\tsc{DE}

\begin{document}
\let\WriteBookmarks\relax
\def\floatpagepagefraction{1}
\def\textpagefraction{.001}
\shorttitle{Mechanical Properties of hierarchical schwarzites-based structures}
\shortauthors{Bastos V. Leonardo et~al.}

\title [mode = title]{Mechanical Energy Absorption of Architecturally Interlocked Petal-Schwarzites  }

\author[1]{Leonardo V. Bastos}
\author[2]{Rushikesh S. Ambekar}
\author[2]{Chandra S. Tiwary}
\author[3,4]{Douglas S. Galvao}
\cormark[2]
\ead{galvao@ifi.unicamp.br}
\author[1]{Cristiano F. Woellner}[orcid=0000-0002-0022-1319]
\cormark[1]
\ead{woellner@ufpr.br}

\address[1]{Physics Department, Federal University of Parana, Curitiba-PR, 81531-980, Brazil}
\address[2]{Department of Metallurgical and Materials Engineering, Indian Institute of Technology Kharagpur, West Bengal, India, 721302}
\address[3]{Applied Physics Department, Gleb Wataghin Institute of Physics, State University of Campinas, Campinas,SP, 13083-970, Brazil}
\address[4]{Center for Computing in Engineering \& Sciences, State University of Campinas, Campinas, SP, 13083-970, Brazil}


\begin{abstract}
We carried out fully atomistic reactive molecular dynamics simulations to study the mechanical behavior of six newly proposed hybrid schwarzite-based structures (interlocked petal-schwarzites). Schwarzites are carbon crystalline nanostructures with negative Gaussian curvature created by mapping a TPMS (Triply Periodic Minimal Surface) with carbon rings containing six to eight atoms. Our simulations have shown that petal-schwarzite structures can withstand uni-axial compressive stress up to the order of GPa and can be compressed past 50 percent strain without structural collapse. Our most resistant hierarchical structure has a calculated compressive strength of 260~GPa and specific energy absorption (SEA) of 45.95 MJ/kg, while possessing a mass density of only 685 kg/m$^3$. These results show that these structures could be excellent lightweight materials for applications that require mechanical energy absorption.
\end{abstract}




\begin{keywords}
Molecular Dynamics \sep Stress-Strain  \sep Energy absorption \sep Schwarzite \sep Interlocked Structure
\end{keywords}

\maketitle

\doublespacing

\section{Introduction}

Schwarzites are crystalline 3D carbon allotrope structures proposed in 1991 by Mackay and Terrones \cite{mackay_1991,mackay-terrones,terrones_1992,Lenosky1992}. Schwarzites were named after the German mathematician Hermann Schwarz and his work on triply periodic minimal surfaces (TPMS). The schwarzites are created by mapping/placing carbon rings containing 6, 7, or 8 atoms into TPMS creating atomic structures with negative Gaussian curvatures. Schwarzites have high porosity and complex topologies, which results in a high surface-to-volume ratio. These structures show unique physical and chemical properties, which were exploited for several applications in mechanical resistant structures, sensing, and water treatments \cite{D1RA03097C,KUMBHAKAR2021126383}.  Recently, macroscale 3D printed Schwarzites revealed a unique mechanical deformation behavior \cite{woellner_2018,mackay-terrones,terrones_1997,phillips_1992,miller_2016,braun_2018,felix_2019,feng_2020}, such as a strong resilience to compressive stress and the ability to withstand extreme structural deformations without fracture. The experimental observations clearly demonstrated a good correlation between topology and mechanical properties, which seem to be scale-independent \cite{felix_2019}. The complex topologies result in unique stress and strain distributions, which is not observed in more conventional geometrical shapes (such as tube, sheets, etc.). By better understanding the structure/topology-properties correlations, we can further improve or build novel topologies/architectures with enhanced mechanical properties.   Recently, these structures were utilized as building blocks to create complex hierarchical architecture with enhanced mechanical properties. These ideas were extended \cite{KYOTANI2000269,ODKHUU201439,TAGAMI2014266,ZHANG2018289,OLIVEIRA2018782,PEDRIELLI2017796,eliezer2022,OLIVEIRA2019190, buehler2023} to build several unique architectures with exceptional mechanical properties \cite{douglas2022pentadiamond, 3Ddouglas2022, ambekar2021flexure, sajadi20193d}. 

In general, the rigidity of porous architecture is limited by the solid versus void content ratio. For example, the stiffness of carbon nanotube improves as we increase the number of walls.  The internal wall interactions contribute to improving the mechanical resistance under different loading, but the sliding of walls reduces the elastic modulus values. Inspired by these examples of multi-walled/layered structures, in this work, we have used different schwarzites' unit cells from the primitive family; some of them are shown in Figure \ref{fig:dolls}($\alpha=1$). In Table \ref{tab:tabela-info}, we present the structural information of these schwarzites. 

\begin{figure}
\centering
\includegraphics[scale=0.3]{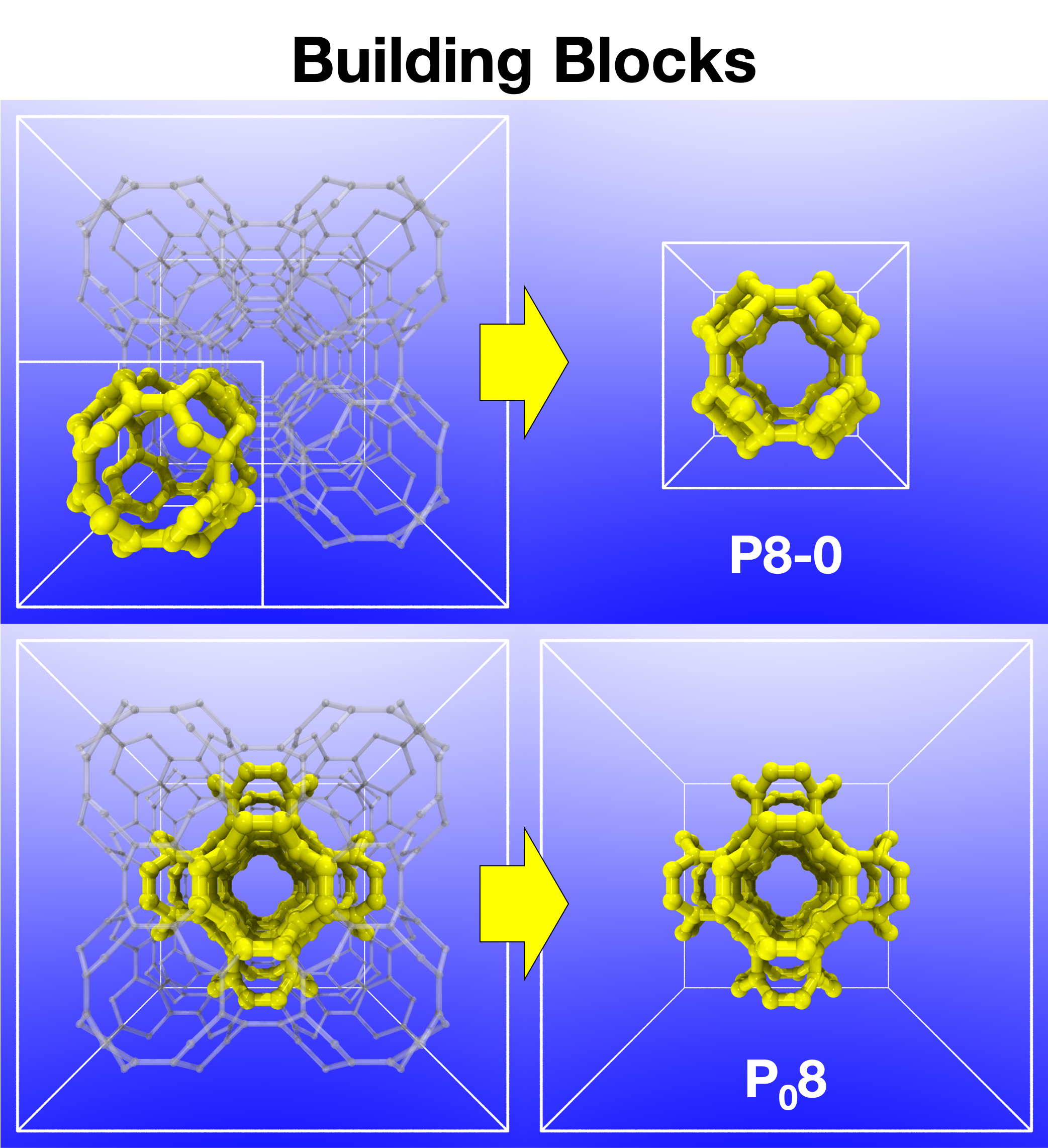}
    \caption{The hybrid structures are composed of multiple schwarzites building blocks (top part). In order to create  the interlocked petal-schwarzites (bottom), the original Schwarzites' unit cells are modified in order to allow a perfect match with the next unit cell layers. The cases of P8-0 and P08 schwarzites are shown here. See text for discussions.}
    \label{fig:Building-Blocks}
   \end{figure}

\begin{figure}
\centering
\includegraphics[scale=0.22]{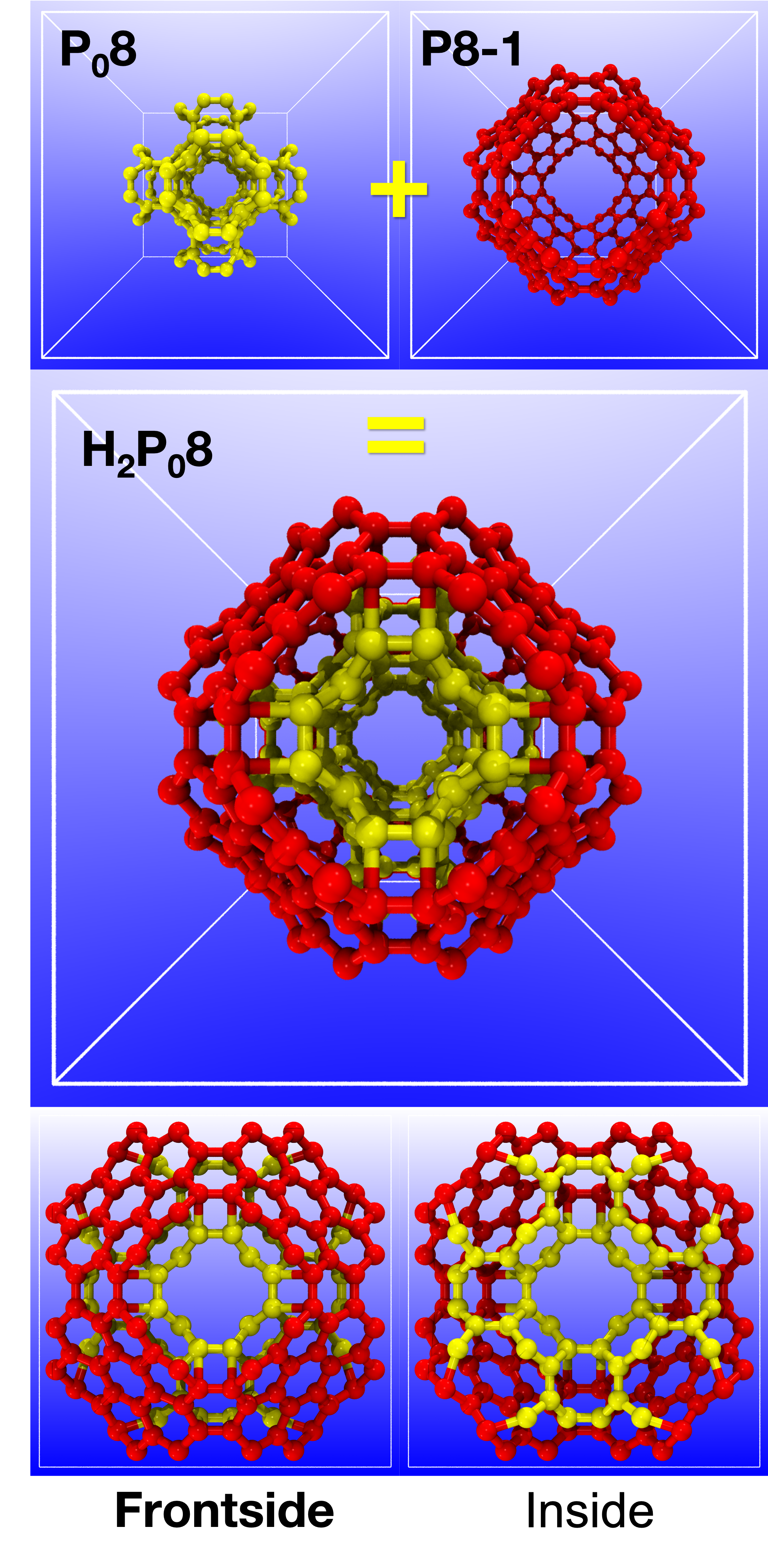}
    \caption{Composition of the interlocked petal-schwarzite $H_{2}P_{0}8$. Its internal layer corresponds to the $P_{0}8$ unit cell, which is illustrated in Figure \ref{fig:Building-Blocks}, while the external layer comes from schwarzite P8-1. See text for discussions.}
    \label{fig:H1P081}
\end{figure}

\begin{figure}
\centering
\includegraphics[scale=0.35]{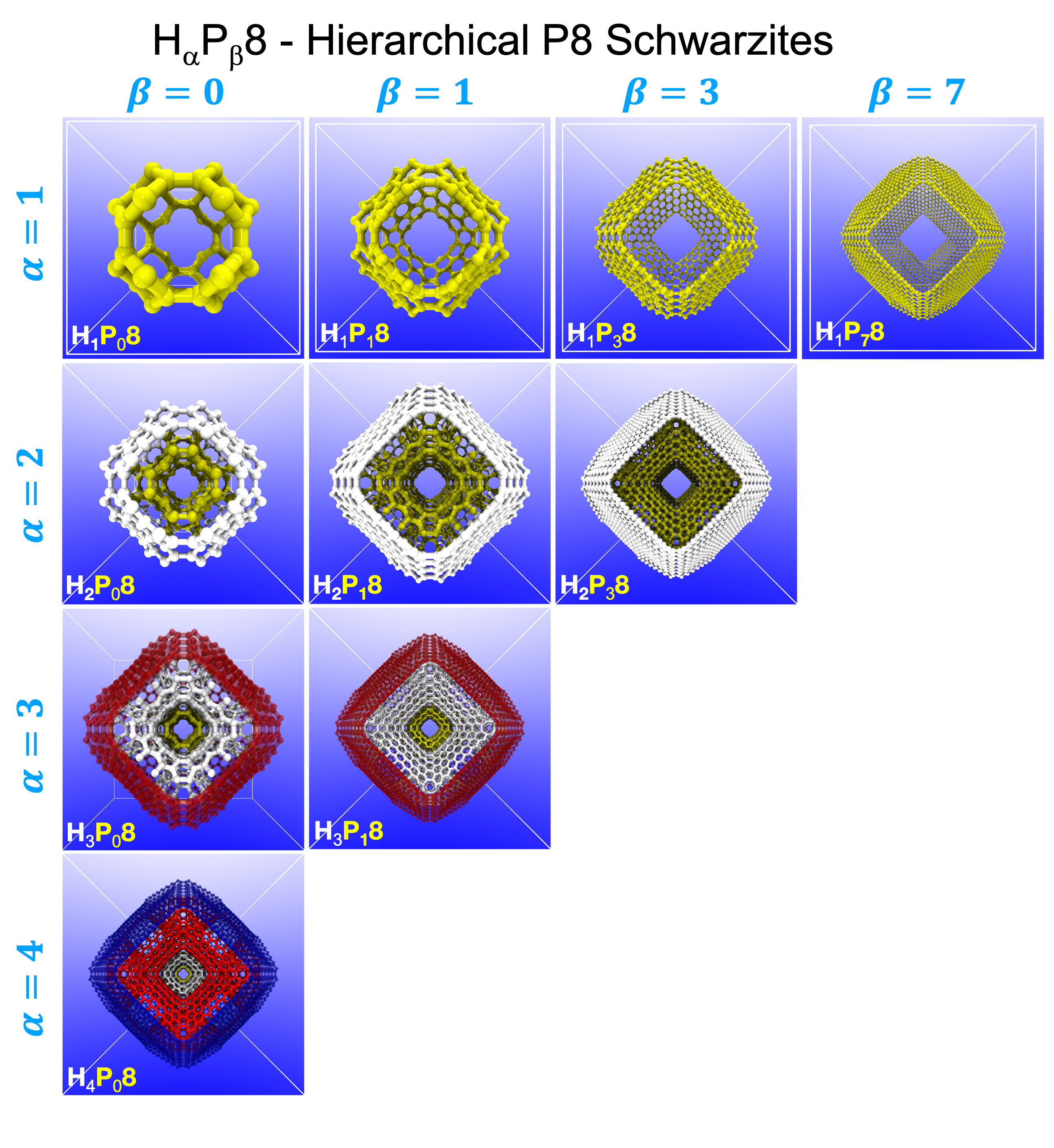}
    \caption{The petal-schwarzites structures investigated in the present work, along with the system adopted to label them. The $\alpha$ value indicates the number of layers each structure possesses, while $\beta$ indicates the smallest possible schwarzite present in its arrangements. The $H_{4}P_{0}8$ structure, for example, has four layers, which means it contains all our four schwarzites, and the $H_{1}P_{0}8$ one, in the form of the $P_{8}0$ unit cell, is the smallest structure present in it. Each layer was assigned a different color to better illustrate the concept.}
    \label{fig:dolls}
\end{figure}

As the Figure \ref{fig:Building-Blocks} and Table \ref{tab:tabela-info} show, the structures have different lattice sizes. With this in mind, we propose combining the structurally different schwarzites to create new hybrid multi-layered interlocked petal-schwarzites. The building process is described in Figures \ref{fig:Building-Blocks}-\ref{fig:H1P081}, and it was used to create six novel petal-schwarzite structures. This process aims to create structures that enhance some of the schwarzite mechanical properties, with potential emphasis on applications as mechanical energy absorbers.

We carried out fully atomistic molecular dynamics simulations to study the mechanical behavior of all our pure schwarzites and the petal ones under compressive stress. We have, in particular, investigated how their different topologies (such as how the ratio of hexagons/octagons, which impacts the curvature) affect their mechanical deformations and fracture pattern. We also calculated the mechanical energy absorbed during the compressive process.

\section{Methodology}

We have analyzed the mechanical properties of four schwarzites from the Primitive family \cite{mackay-terrones} and six related petal-schwarzites structures through fully atomistic reactive molecular dynamics simulations. All calculations were carried out using the Adaptive Intermolecular Reactive  Border (AIREBO-M) \cite{AIREBO-M} force field, an empirical potential parameterized to model hydrocarbon structures. We used it as implemented in the Large-scale Atomic/Molecular Massively Parallel Simulator (LAMMPS) \cite{LAMMPS}. AIREBO-M is a reactive force field, which allows modeling the formation/breaking of chemical bonds within the system, and it can be used to treat large-scale simulations being much less computer time-consuming than fully-based quantum methods. 

\begin{table}
\caption{Information about all the structures investigated here. "n" is the number of atoms in a unit cell, "a" is the structures' lattice parameter, and "$\rho$" is their mass density.}
\label{tab:tabela-info}
\begin{center}
\begin{tabular}{|c|c|c|c|}
\hline 
Structure & n    & a(\AA)  & $\rho(g/cm^{3})$   \\
\hline
$H_{1}P_{0}8$     & 48    &  7.80    &  2.017 \\ 
$H_{1}P_{1}8$     & 192   &  14.91   &  1.155 \\ 
$H_{1}P_{3}8$     & 768   &  29.02   &  0.626 \\ 
$H_{1}P_{7}8$     & 3072  &  56.92   &  0.332 \\ 
\hline
$H_{2}P_{0}8$    & 312   &  14.91   &  1.876 \\ 
$H_{3}P_{0}8$   & 1416  &  29.02   &  1.155 \\ 
$H_{2}P_{1}8$    & 1296  &  29.02   &  1.057 \\ 
$H_{2}P_{3}8$   & 5688  &  56.92   &  0.615 \\ 
$H_{3}P_{1}8$   & 6216  &  56.92   &  0.672 \\ 
$H_{4}P_{0}8$  & 6336  &  56.92   &  0.685 \\ 
\hline
\end{tabular}
\end{center}
\end{table}

Our study was carried out in two parts. First, we performed a slow uni-axial compressive test at a constant rate. The simulations for this part were carried out with an NPT ensemble and using a 0.25~fs timestep and with periodic boundary conditions along all directions. Before compressing the structures, a thermalization process was performed for 5~ps, split into two parts of equal duration. In the first half, the pressure was set to zero along all three directions, and during the second half, the pressure along the z-direction, the direction in which the compression would later be applied, was allowed to change. For all structures, it was tested whether the stress had reached a stationary and close to zero value by the end of the thermalization process and also whether their potential energy stabilized. These two tests ensure that thermal effects will not produce spurious effects in the stress analysis during the compression process.

After the thermalization process, the structures are then compressed along the z-direction at a constant strain rate of $10^{-2}$ps$^{-1}$ for 100 ps. The structure's mechanical properties were studied during this part of the simulation by analyzing the stress-strain curve for each one of them. In this context, the strain is defined as:

\begin{equation}
    \epsilon_{L} = \frac{|L - L_{0}|}{L_{0}},
\end{equation}

where $L_{0}$ is the structure's initial length and $L$ is the deformed structure's length.
The stress distribution was analyzed by calculating the pressure tensor\cite{LAMMPS_stress}, which is a symmetric tensor of second order whose components are given by the relation:

\begin{equation}
    \sigma_{ij} = \sum_{k}^{N}\frac{m_{k}v_{ki}v_{kj}}{V} + \sum_{k}^{N'}\frac{r_{ki}f_{kj}}{V},
\end{equation}

where $N$ is the number of atoms in the system, $N'$ is the number of atoms, including periodic images, and $V$ is the system volume. The first term is the kinetic energy, and the second, the virial term, is a sum of pair, bond, angle, dihedral, improper, long-range, and fixed contributions to the force on each atom. For our analysis, we need this tensor's diagonal component along the direction of deformation, which will be used as the compressive stress in the stress-strain plot. From this tensor component for a single atom, we also calculated the von Mises stress for each atom in the system, which is defined as:

\begin{equation}
\sigma = \sqrt{\frac{(\sigma_{xx} - \sigma_{yy})^{2} + (\sigma_{xx} - \sigma_{zz})^{2} + (\sigma_{yy} - \sigma_{zz})^{2} +6(\sigma_{xy}^{2} + \sigma_{xz}^{2} + \sigma_{yz}^{2})}{2}},
\end{equation}

where $\sigma_{ij}$ are the components of the pressure tensor. Since the von Mises stress is calculated for every atom and not as a global property of the system, it was used to analyze the spatial distribution of stress during our simulations. The first calculated quantity is the structures' Young's Modulus values, which quantifies its elasticity. It is estimated from the linear portion of the stress-strain curve and is defined as the angular coefficient relating stress and strain, as given by:

\begin{equation}
    \sigma_{L} = E \epsilon_{L},
\end{equation}

where $\sigma_{L}$ is the pressure component along the compression direction,  $\epsilon_{L}$ is
the strain and $E$ is Young’s modulus. In this linear region, we also define the structures' yield strength, which is the value of stress that indicates the end of the linear regime. 


The final mechanical characteristic analyzed was the structure's energy absorption, which can also be calculated from the stress-strain curve, as given by:

\begin{equation}
    \int_{0}^{\epsilon} \sigma(\epsilon') \,d\epsilon',
\end{equation}

where $\sigma(\epsilon')$ is the stress, and $\epsilon$ is the strain. Again, for better comparison, we divided the energy absorption by the structure's mass densities, thus obtaining the specific energy absorption, and used the fracture strain that will be determined as the integral limit.

As can be seen from Table \ref{tab:tabela-info}, our structures have very different lattice sizes, and thus the supercells generated from them will also vary in size. In order to warrant that the results reported here are actually related to the difference in topologies and are not merely the result of the size differences, we also performed our tests on supercells of different sizes to ensure that the conclusions were size independent.


\section{Results}

Supercells with sizes between the primitive cell and a 5x5x5 were created after replicating the unit cell of the same structure. The simulated mechanical properties and the stress-strain curve behavior for supercell sizes are similar, as shown in Figure S1. Thus, a 2x2x2 supercell is large enough to represent the properties of our structures, as can be seen in Figure S1 for the schwarzite $H_{1}P_{0}8$.

\begin{figure}
\centering
\includegraphics[scale=0.5]{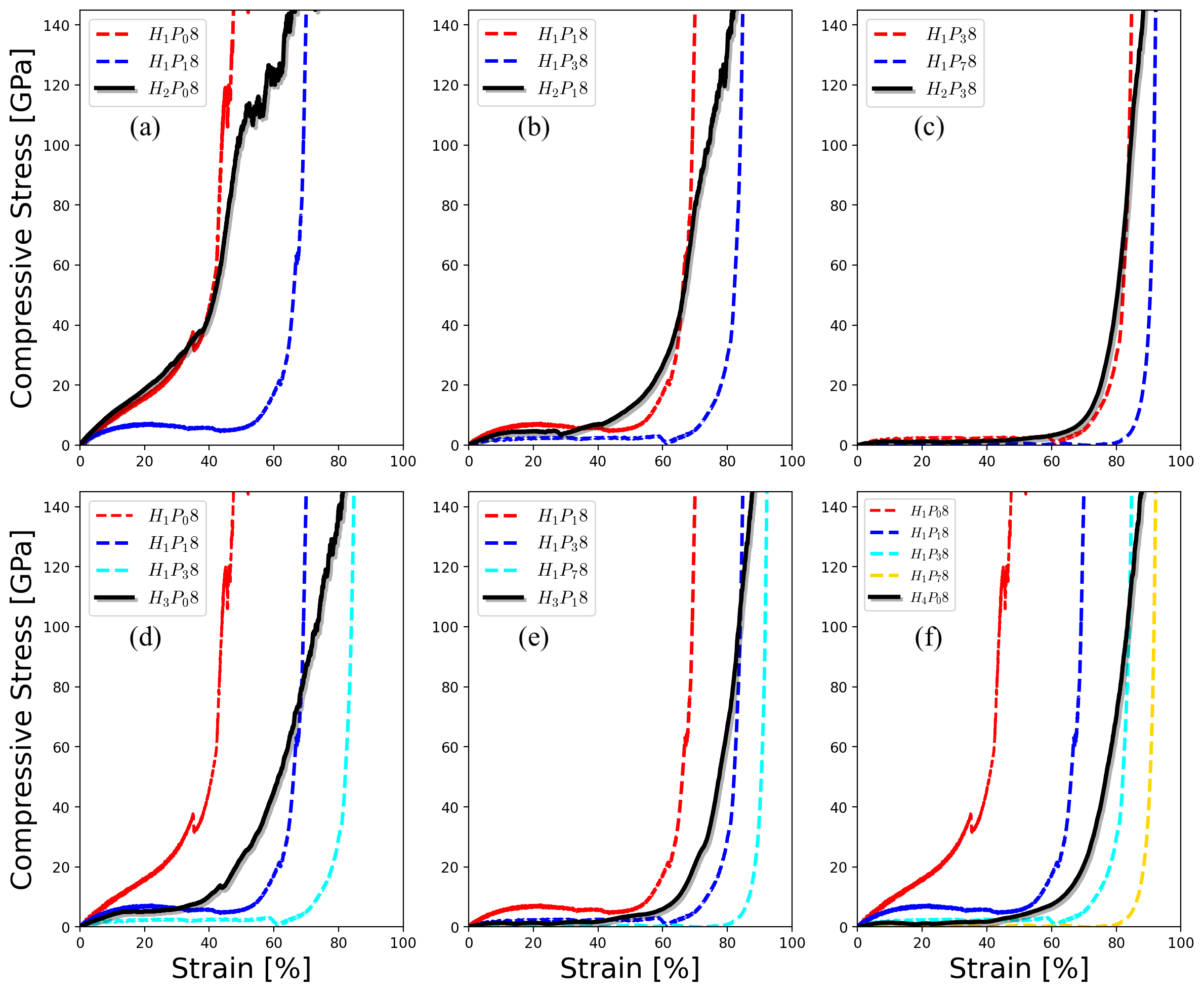}
    \caption{Stress-strain curves for the hybrid structures. Each plot compares a hybrid structure to all the individual schwarzites used to create it. We can see that the hybrid structure densification regimes start earlier and last longer in comparison to the strongest schwarzites that contain them, as both were found to have very similar fracture strains. This feature results in much higher specific energy absorption values.}
    \label{fig:stress-strain-dolls}
\end{figure}

The stress-strain curves for all the schwarzites are shown in Figure \ref{fig:stress-strain-dolls} and in supporting information (Figure S2). Our stress-strain curves (except $H_{1}P_{0}8$) present behaviors that are typical of foam-like materials \cite{stress-strain-foams}, whose curves are characterized by a linear rise in compressive stress for low deformations until the yield strength is reached. Then is followed by a plateau that lasts until the densification regime, when there is a sharp increase in stress until the structure finally collapses/breaks. Carbon-based foams have already been experimentally demonstrated to be able to be compressed without fracture to extreme deformation levels. Due to their foam-like structure, schwarzites also share this high resistance to compression and can be compressed up to more than 80\% strain before collapsing, as can be inferred from their stress-strain curves \cite{woellner_2018}.

For the four schwarzites indicated in  Table \ref{tab:tabela-stress-strain}, which contains the values for the quantities estimated from the stress-strain curves, we can see that Young's modulus (E) values, the specific energy absorption (SEA), and the resilience to compressive stress are higher for the structures with a higher ratio of hexagons to octagon carbon rings. The $H_{1}P_{7}8$, which has the highest ratio, was found to be the most deformable (it is the lowest dense structure) and can maintain its structural integrity up to 70\% deformation while accumulating very low levels of compressive stress. A comparative stress-strain curve for interlocked petal-schwarzites with the outermost and innermost layers is shown in Figure \ref{fig:stress-strain-dolls}. 

In the earlier stages of compression, the petal-schwarzites (H$_2$P$_0$8) stress-strain curve follows that of the innermost layer (P08) (Figure \ref{fig:stress-strain-dolls}a). It shows that the petal-schwarzite initial deformation follows the deformation similar to the inner (P08) schwarzites' architecture until it reaches yielding. As we deform the petal-schwarzite more than 40\%, the curve approaches the path of the outer layer (P8-1) with few serrations. 

It is important to note that the inner layer does not have a broad plateau region, whereas the outer layer has a low-yielding but large plateau region. The petal-schwarzite shows a combination of both, i.e., higher yield and a densification region with high stress. Thus, we can see that the petal-schwarzite has the best of both the inner and outer layers.  Figure 4b shows different stress-strain behavior for petal-schwarzites (H$_2$P$_1$8) compared to its inner (H$_1$P$_1$8) and outer (H$_1$P$_3$8) layers. The yielding of petal-schwarzites is lower than the inner layer, but the onset of densification starts earlier than the inner layer. As soon as we cross the 60\% deformation, it starts following the path of the outer layer (H$_1$P$_3$8). In Figure \ref{fig:dolls}c, we see a different trend, where the petal-schwarzite's (H$_2$P$_3$8) curve matches that of the innermost structure (H$_1$P$_3$8) during most of the compression. Both curves enter the densification regime at the same strain (about 60\%). However, at the later stages of deformation, we see that the petal-schwarzite's curve follows the path of the outermost layer (H1P78), reaching the same final value of compressive strength. Again, the petal-schwarzite is shown to maintain the best properties of the inner (higher yield and compaction) and outer layer (high strength for high compaction). 

Figures \ref{fig:stress-strain-dolls}d-e, f represent the three-layered structures. We see a common behavior in these two petal-schwarzites: the petal-schwarzites curve (H$_3$P$_0$8 and H$_3$P$_1$8) follows the second innermost (H$_1$P$_1$8 and H$_1$P$_3$8) structure's curve. However, in Figure 4d, the petal-schwarzite enters the densification regime much earlier, approaching the slope of the innermost curve. This indicates that the inner layer has a stronger effect on the behavior of the petal-schwarzite H3P08. In Figure 4e, we notice that the petal-schwarzite's curve follows the second innermost curve until this layer collapses, which can be seen in Figure 4e as a slight drop at about 60\% strain. In both cases, we verify again that the curves approach those of the outermost layer, which means they maintain the compressive strength of their most resistant layer. Figure 4f shows the curve for the 4-layered petal-schwarzites (H$_4$P$_0$8). We see that its curve follows the second outermost layer (H1P18) for most of the compression process. It is important to note that while the curve for the inner layer shows a collapse at around 60\% deformation, the curve for the petal-schwarzite still shows resistance against loading.  

In summary, from Figure \ref{fig:stress-strain-dolls}, we can conclude that the most resistant structure is able to withstand deformations of up to 90\% strain, despite being the lowest dense one. For the petal-schwarzites structures, we found that they were virtually equivalent to the strongest schwarzite that composes them regarding the fracture strains and compressive strengths while having higher values of specific energy absorption. These results show that their properties are comparable with high-performance energy-absorbing materials \cite{sajadi_2019_3D_tubulanes}. 

\begin{figure}
    \centering
    \includegraphics[scale=0.8]{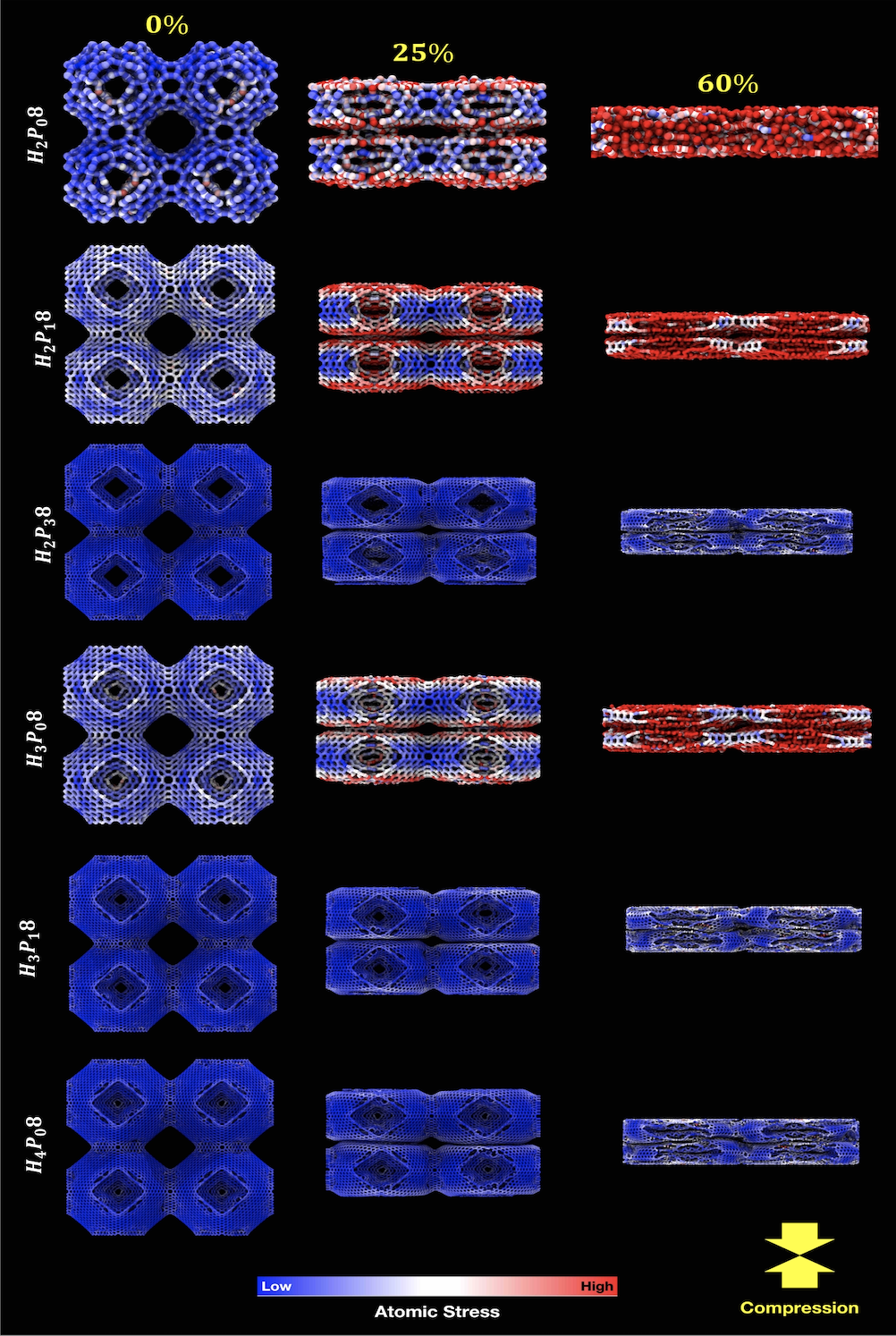}
    \caption{Stress distribution of six different petal-schwarzites under three different strain (0, 25, and 60\%) regimes. The color scale bar shows the high stress for red and low for blue, respectively.}
    \label{fig:snap-stress-all}
\end{figure}

Figure \ref{fig:snap-stress-all} shows the snapshots of von Mises stress distribution of the six petal-schwarzites under three different strain (0, 25, and 60\% strain) regimes (supporting videos SV1 and SV2). Following the von Mises distribution, We divided the structures into two groups. The stress distribution of group-I (H$_2$P$_0$8, H$_3$P$_0$8, H$_2$P$_1$8) is very different, as compared to group-II (H$_4$P$_0$8, H$_3$P$_1$8, H$_2$P$_3$8). At 25\% strain, we can clearly observe very low-stress concentration in group-II structures. On the other hand, the group-I structures show high and low-stress concentrations. The high-stress concentrations are observed along the loading direction, whereas the low-stress concentrations are observed perpendicular to the loading direction. 

At 60\% strain, the group-I structures show more deformation (densification) as compared to group-I. Also, the stress accumulation is higher in group-I than in group-II structures. The H$_2$P$_0$8 structures show the highest uniform stress accumulation over H$_3$P$_0$8 and H$_2$P$_1$8. The group-II structures show lower stress accumulation perpendicular to the loading direction. Based on the snapshot, we can clearly conclude that as we increase the number of interlocked layers, the mechanical resistance increases. Also, as we increase the number of layers, we also observe more energy absorption. Thus, it is important to understand the deformation behavior/stress accumulation across the interlocked layers of petal-schwarzites. 

\begin{figure}
    \centering
    \includegraphics[scale=0.04]{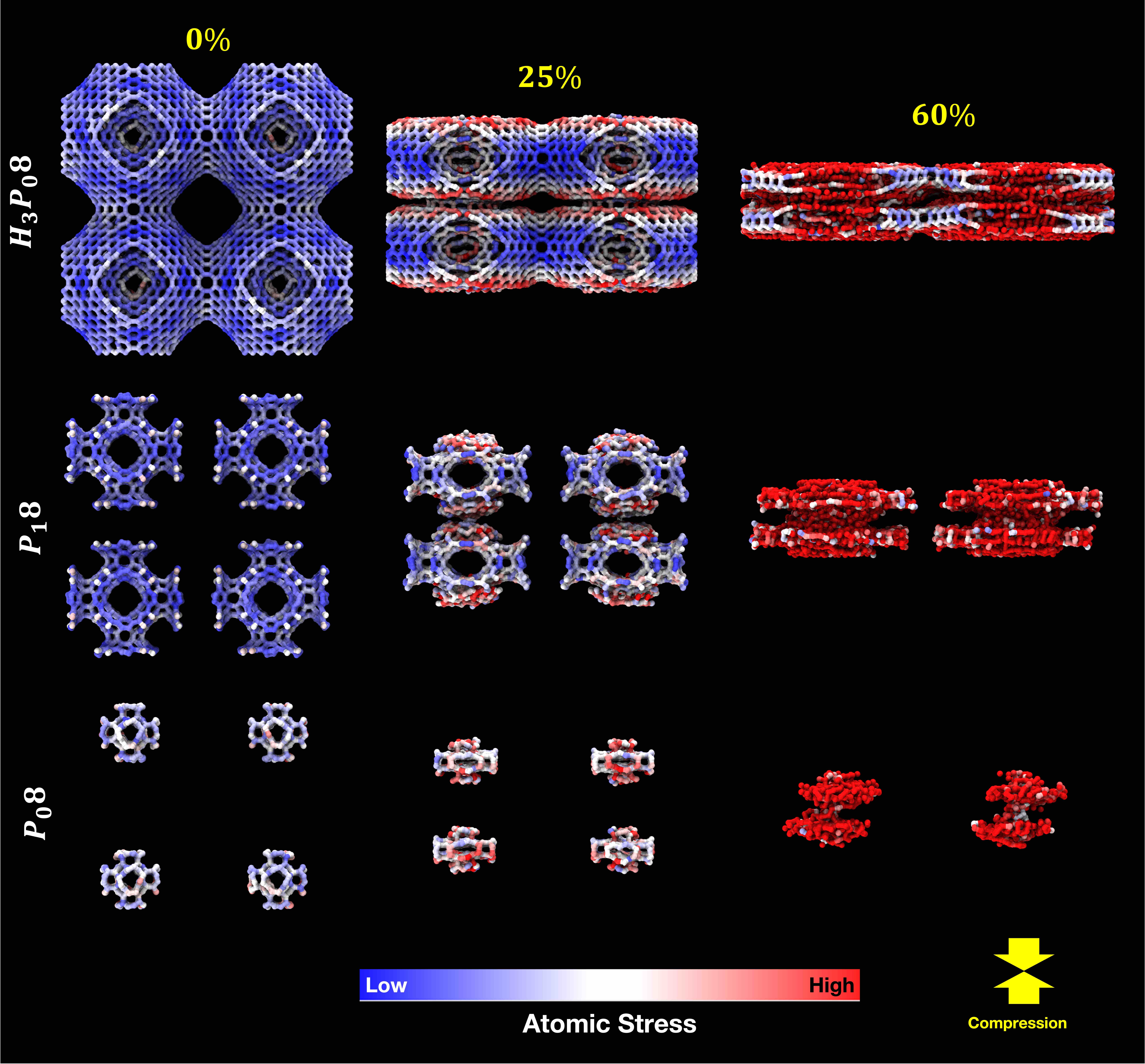}
    \caption{Stress distribution in different layers of the H3P08 petal-schwarzites under three different stain regimes. The color scale bar shows the high red and low blue stress, respectively.}
    \label{fig:snap-stress-3Ls}
\end{figure}

Figure \ref{fig:snap-stress-3Ls} shows the stress distribution across the different layers of H$_3$P$_0$8 petal-schwarzite under different strain regimes. Notice that at 25\% strain, the inner layer is responsible for accumulating most of the stress (which is accumulated along the direction of compression, as can be seen in the Figure \ref{fig:snap-stress-3Ls}). In contrast, the outer layers have more evenly distributed stress patterns. This explains why the stress-strain curves of the petal-schwarzites tend to follow the curve of the innermost schwarzites up to the point where the inner layers start to collapse (this behavior is shown in Figure \ref{fig:stress-strain-dolls}). For 60\% strain, we can see that the innermost layer has already collapsed, resulting that the other layers start supporting the load. At this point, notice that the second innermost layer is where the stress accumulates, as areas with low-stress accumulation can still be seen in the outer layer. For all petal-schwarzites, we have observed the same tendency where the inner layers accumulate more stress and, as a result, fail earlier than the outer ones.

\subsection{Energy absorption results}

\begin{figure}
    \centering
    \includegraphics[scale=0.6]{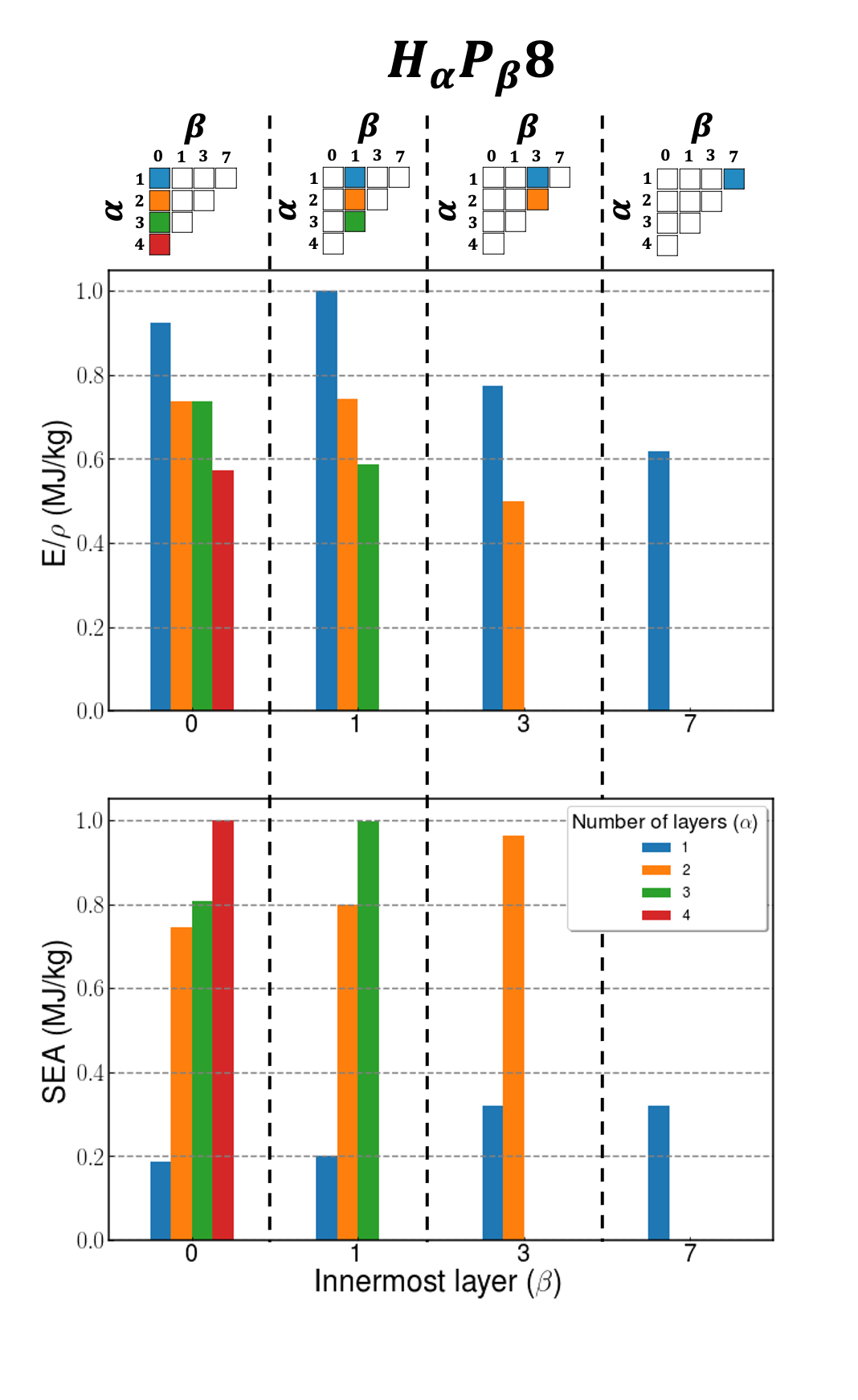}
    \caption{Specific modulus (E/$\rho$) and specific energy absorption (SEA) for all structures (pure and hybrid ones). Among hybrid structures with the same innermost layer that possess the same value of $\beta$, it can be seen that an increase in the number of layers leads to higher values of SEA and lower values for Young’s modulus, which means that structures with more layers are less elastic. However, the decrease in elasticity is much less significant if compared to the SEA increase, as four-layered structures, for example, possess SEA values four times as large as the ones found for a single-layered one. In contrast, their normalized Young’s modulus values are more than half of the single-layered structures.}
    \label{fig:graf_barra}
\end{figure}

In Figure \ref{fig:graf_barra}, we compare the effect of the number of layers ($\alpha$) of all schwarzites on the specific modulus (E/$\rho$) and SEA. Regarding their elasticity, it was found that in a four-layered petal-schwarzite structure, Young's modulus is typically half of the value obtained for the single-layered structure (schwarzite). As the highest Young's modulus value was estimated to be in the order of $10^{2}$~GPa, all our structures are much more elastic than typical crystalline carbon allotropes, reaching values that are roughly 10\% of graphene and diamond. A general trend was also observed among petal-schwarzite structures of the same type (which possess the same value of $\beta$ — refer to Figure \ref{fig:dolls} for an explanation of the methodology adopted to name these structures), in which an increase in the number of layers always represents an increase in the structure Specific Energy Absorption (SEA).

These results imply that the petal-schwarzite structures can absorb more energy than the pure schwarzites before collapsing. This favors their choice over any single-layered schwarzite for potential applications requiring lightweight mechanical absorbing materials. Quantitatively, this advantage is reflected by the fact that the calculated SEA value for the strongest hierarchical structure reached a value of 45.95 MJ/kg, which is more than three times larger than the value obtained for the strongest schwarzite. The results for all our structures, which all favor any petal-schwarzite structure over any schwarzite, can be seen in Table \ref{tab:tabela-stress-strain}.  They show that our petal-schwarzite structures could be excellent lightweight candidates for applications that require significant mechanical energy absorption.


\begin{table}
\centering
\begin{tabular}{ |c|c|c|c|c| }
\hline 
Structure & E (Gpa) & E/$\rho$ (MJ/kg) & SEA (MJ/kg) & $\epsilon_{F}$    \\
\hline
$H_{1}P_{0}8$     & 116.06  & 57.54 &  8.59    & 0.50     \\
$H_{1}P_{1}8$     & 71.86   & 62.22 &  9.20    & 0.71     \\
$H_{1}P_{3}8$     & 30.17   & 48.19 &  14.68   & 0.86     \\  
$H_{1}P_{7}8$     & 12.78   & 38.49 &  14.71   & 0.93    \\ 
\hline
$H_{2}P_{0}8$    & 86.21    & 45.95 &  34.18   & 0.84     \\ 
$H_{2}P_{1}8$    & 48.85    & 46.22 &  36.73   & 0.91       \\     
$H_{2}P_{3}8$    & 19.14    & 31.12 &  44.22   & 0.95       \\     
$H_{3}P_{0}8$    & 53.05    & 45.93 &  37.12   & 0.91       \\ 
$H_{3}P_{1}8$    & 24.52    & 36.49 &  45.87   & 0.96        \\ 
$H_{4}P_{0}8$    & 24.43     & 35.67 &  45.95   & 0.96     \\
\hline
\end{tabular}
\caption{Summary of the mechanical properties. Here, E is Young's modulus, E/$\rho$ is the specific modulus, SEA is the specific energy absorption, and $\epsilon_{F}$ is the fracture strain. Notice that the schwarzites $H_{1}P_{0}8$ and $H_{1}P_{1}8$, which are denser than the schwarzites $H_{1}P_{3}8$ and $H_{1}P_{7}8$, have much inferior performance on the tests. The same trend is observed among the hybrid structures, as the denser ones were found to be less resistant.}
\label{tab:tabela-stress-strain}
\end{table}

\pagebreak

\section{Summary and Conclusions}

We carried out fully atomistic molecular dynamics simulations to study the mechanical behavior of schwarzite-based structures. Based on schwarzites from the primitive (P) family, six new interlocked petal-schwarzite (hybrid) structures are proposed. Similar to the pure schwarzites, these new structures possess very low density (as low as 0.685 g/cm$^3$) while showing a very high capacity for mechanical energy absorption.

Based on the stress-strain curves for our structures, we estimated that our most resistant petal-schwarzite structure is able to withstand deformations of up to 90\% strain, despite being the lowest dense one. This resistance to deformation is quite similar to the one observed for foam-like materials, which typically present great capacity to be compressed. 

Regarding their elasticity, it was found that in a four-layered petal-schwarzite structure, Young's modulus is typically half of the value obtained for the single-layered structure (schwarzite). As the highest Young's modulus value was estimated to be in the order of $10^{2}$ GPa, all our structures are much more elastic than typical crystalline carbon allotropes, reaching values that are roughly 10\% of graphene and diamond. A general trend was also observed among petal-schwarzite structures of the same type; an increase in the number of layers always represents an increase in the structure's specific energy absorption (SEA).

These results show that the petal-schwarzite structures can absorb more energy than the pure schwarzites before collapsing. This favors their choice over any single-layered schwarzite for potential applications requiring lightweight mechanical absorbing materials. Quantitatively, this advantage is reflected by the fact that the calculated value of SEA for the strongest hierarchical structure reached a value of 45.95 MJ/kg, which is more than three times larger than the value obtained for the strongest schwarzite. This show that our petal-schwarzite  structures can be excellent lightweight candidates for applications that require significant mechanical energy absorption. The current approach can be used to build any family and any number of layers of multi-layered interlocked petal-schwarzite architectures. 

\clearpage 
\section*{Supporting Information}
The supplementary material has additional data regarding electronic band structures and the calculation of the elastic modulus from molecular dynamics.

\setcounter{figure}{0}
\renewcommand{\thefigure}{S\arabic{figure}}
\begin{figure}
    \centering
    \includegraphics[scale=1.0]{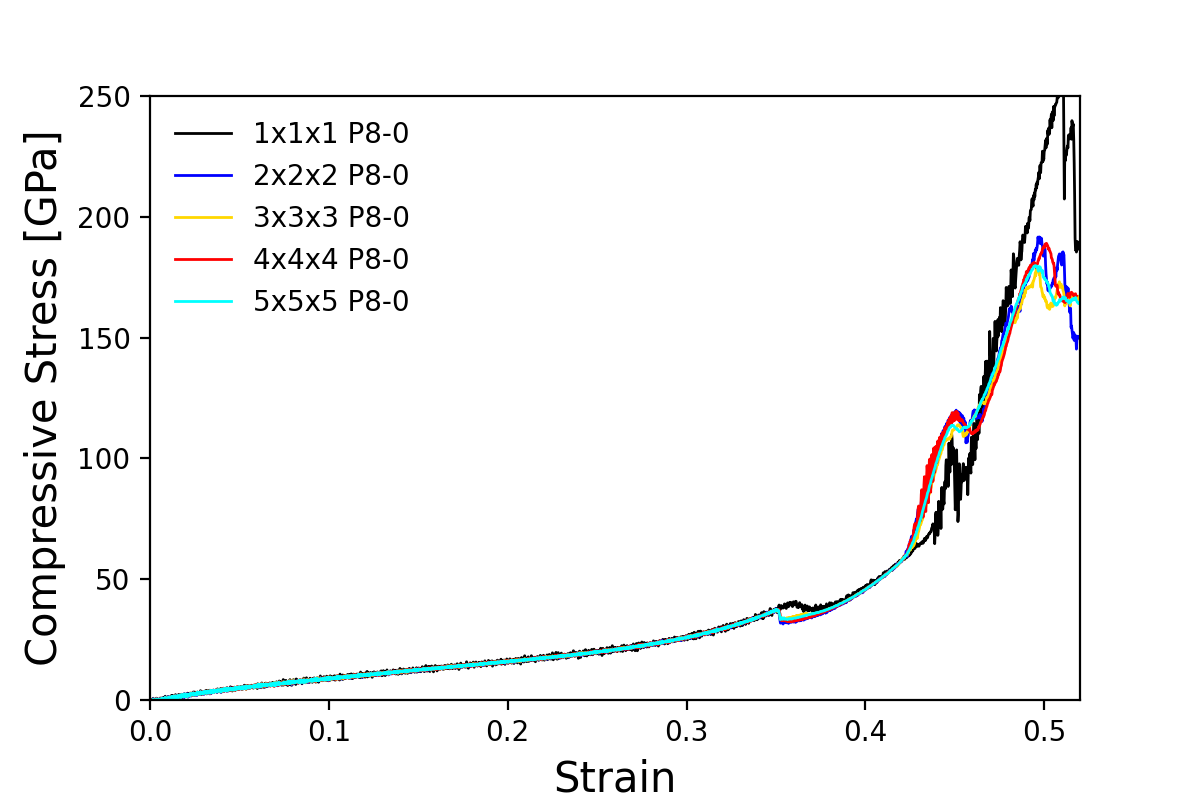}
    \caption{Stress-strain curves for the P8-0 schwarzite structure comparing different  supercell sizes between the primitive cell and a 5x5x5.}
    \label{fig:P80-size-comparison}
\end{figure}

\begin{figure}
    \centering
    \includegraphics[scale=1.0]{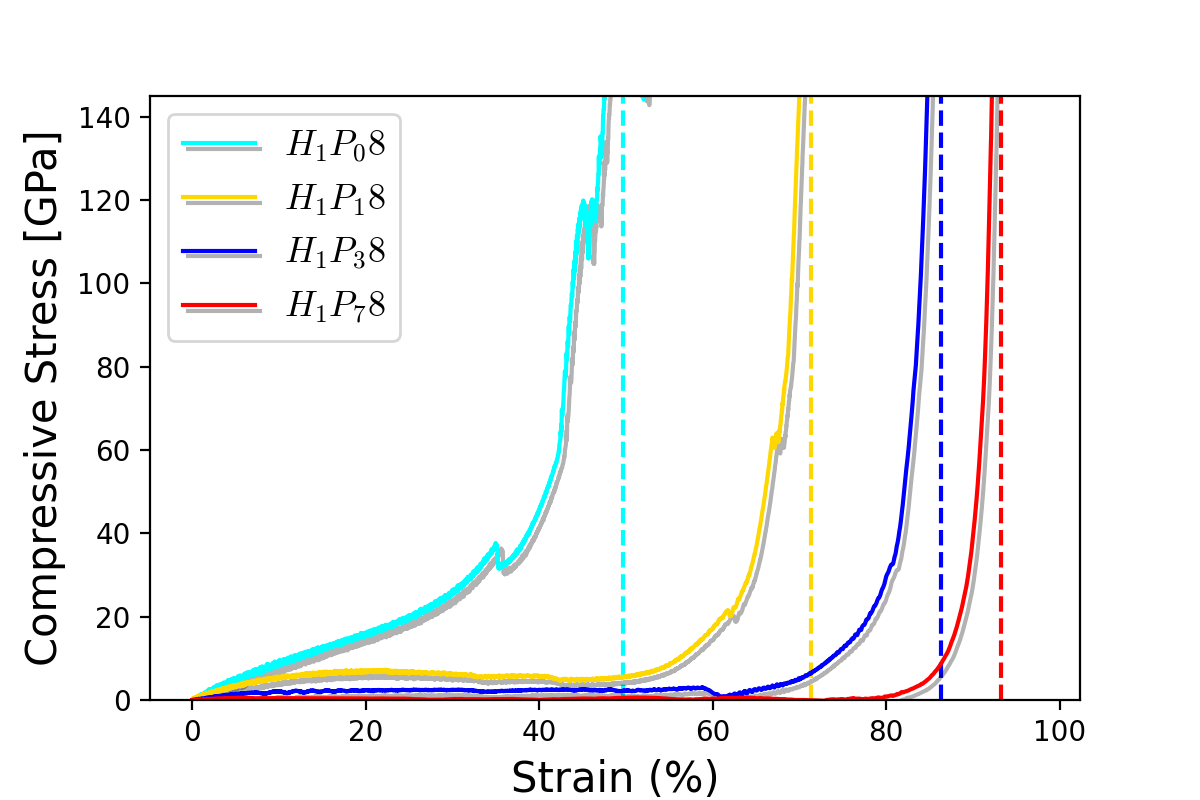}
    \caption{Stress-strain curves for the P8-0 schwarzite Structure. The dotted lines indicate the fracture strain for the structures, and their compressive strength value is defined as the stress value at this breaking point.}
    \label{fig:S2-Schwarzitas-SS}
\end{figure}

\clearpage 
\section*{Data availability statement}

The data that support the findings of this study are available from the corresponding author upon reasonable request.
\clearpage 
\section*{Acknowledgements}
This work was financed in part by the Coordenacão de Aperfeiçoamento de Pessoal de Nível Superior - Brasil (CAPES) - Finance Code 001, CNPq, and FAPESP. We thank the Center for Computing in Engineering and Sciences at Unicamp for financial support through the FAPESP/CEPID Grants \#2013/08293-7 and \#2018/11352-7. We also thank Conselho Nacional de Desenvolvimento Cientifico e Tecnológico (CNPq) for their financial support.
\clearpage 
\bibliographystyle{unsrt}

\bibliography{cas-refs}


\end{document}